\documentclass{aa}
\usepackage{amssymb,amsmath,amsfonts}
\usepackage{graphicx}
\usepackage{color}
\usepackage{ulem}             
\usepackage{multicol}
\usepackage{caption}
\usepackage{pgf,tikz}
\usetikzlibrary{arrows}
\usetikzlibrary{trees}
\usetikzlibrary{snakes}
\usetikzlibrary{shapes}
\usetikzlibrary{matrix}
\usetikzlibrary{calc}
\usetikzlibrary{positioning}
\usetikzlibrary{chains}
\usetikzlibrary{shadows}

\tikzstyle{Point} = [fill, radius=0.08]
\tikzstyle{BigPoint} = [fill, radius=0.13]
\tikzstyle{Leaf} = [color = gray]
\tikzstyle{Line1} = [dashed]
\tikzstyle{Line2} = [dotted, ultra thick]\usepackage{colortbl}

\usepackage{xcolor}

\usepackage{amssymb}
\usepackage{latexsym}
\usepackage{pstricks}
\usepackage{pst-plot}
\usepackage{tabularx}
\usepackage{mathtools}
\usepackage{bm}
\usepackage{soul}
\usepackage{cancel}
\usepackage{hyperref}   

	\usetikzlibrary[patterns]

\newcommand{\be}{\begin{equation}}
\newcommand{\ee}{\end{equation}}

 {\everymath{\displaystyle\everymath{}}\array}%
 {\endarray}

\begin{document}

\title{Alleviating the Transit Timing Variations bias in transit surveys
}
\subtitle{II. RIVERS: Twin resonant Earth-sized planets around Kepler-1972 recovered from Kepler's false positive}
\titlerunning{II. RIVERS: Twin resonant Earth-sized planets around Kepler-1972 recovered from Kepler's false positive}

\author{
A. Leleu$^{1}$, J.-B. Delisle$^{1}$, R. Mardling$^2$, S. Udry$^{1}$, G. Chatel$^3$, Y. Alibert$^{4}$ and P. Eggenberger$^{1}$}
\authorrunning{A. Leleu et al}

\institute{
$^1$ Observatoire de Gen\`eve, Universit\'e de Gen\`eve, Chemin Pegasi, 51, 1290 Versoix, Switzerland.\\
$^2$ School of Physics and Astronomy, Monash University, Victoria 3800, Australia.\\
$^3$ Disaitek, www.disaitek.ai.\\
$^4$ Physikalisches Institut, Universit\"at Bern, Gesellschaftsstr.\ 6, 3012 Bern, Switzerland.\\
}

\abstract
{
Transit Timing Variations (TTVs) can provide useful information for systems observed by transit, by putting constraints on the masses and eccentricities of the observed planets, or even constrain the existence of non-transiting companions. However, TTVs can also prevent the detection of small planets in transit surveys, or bias the recovered planetary and transit parameters. 
 Here we show that Kepler-1972 c, initially the "not transit-like" false positive KOI-3184.02, is an Earth-sized planet whose orbit is perturbed by Kepler-1972 b (initially KOI-3184.01). The pair is locked in a 3:2 Mean-motion resonance, each planet displaying TTVs of more than 6h hours of amplitude over the duration of the Kepler mission. The two planets have similar masses $m_b/m_c =0.956_{-0.051}^{+0.056}$ and radii $R_b=0.802_{-0.041}^{+0.042}R_{Earth}$, $R_c=0.868_{-0.050}^{+0.051}R_{Earth}$, and the whole system, including the inner candidate KOI-3184.03, appear to be coplanar.
  Despite the faintness of the signals (SNR of 1.35 for each transit of Kepler-1972 b and 1.10 for Kepler-1972 c), we recovered the transits of the planets using the RIVERS method, based on the recognition of the tracks of planets in river diagrams using machine learning, and a photo-dynamic fit of the lightcurve.
Recovering the correct ephemerides of the planets is essential to have a complete picture of the observed planetary systems. In particular, we show that in Kepler-1972, not taking into account planet-planet interactions yields an error of $\sim 30\%$ on the radii of planets b and c, in addition to generating in-transit scatter, which leads to mistake KOI3184.02 for a false positive. Alleviating this bias is essential for an unbiased view of Kepler systems, some of the TESS stars, and the upcoming PLATO mission.
}


\maketitle

\section{Introduction}

The most successful technique for detecting exoplanets - in terms of number of planets detected - is the transit method: when a planet passes in front of a star, the flux received from that star decreases. It has been, is, and will be applied by several space missions such as CoRoT, Kepler/K2, TESS, and the upcoming PLATO mission, to try and detect planets in large areas of the sky. When a single planet orbits a single star, its orbit is periodic, which implies that the transit happens at fixed time interval. This constraint
is used to detect planets when their individual transits are too faint with respect to the noise of the data: using algorithms such as Boxed Least Squares \citep[BLS,][]{Kovacs2002}, the data reduction pipelines of the transit survey missions fold each lightcurve over a large number of different periods and look for transits in the folded data \citep[][]{Jenkins2010,Jenkins2016}. This folding of the lightcurve increases the number of observation per phase, hence the signal-to-noise of the transit.

As soon as two or more planets orbit around the same star, their orbits cease to be strictly periodic. In some cases the gravitational interaction of planets can generate relatively short-term Transit Timing Variations (TTVs): transits do not occur at a fixed period any more \citep{Dobrovolskis1996,Agol2005}. The amplitude, frequencies, and overall shape of these TTVs depend on the orbital parameters and masses of the planets involved \citep[see for example][]{Lithwick2012,NeVo2014,AgolDeck2016}. As the planet-planet interactions that generate the TTVs typically occur on timescales longer than the orbital periods, space missions with longer baselines such as Kepler and PLATO are more likely to observe such effects. Since the end of the Kepler mission, several efforts have been made to estimate the TTVs of the Kepler Objects of Interest (KOIs) \citep{Mazeh2013,RoTho2015,Holczer2016,Kane2019}. 

TTVs are a goldmine for our understanding of planetary systems: they can constrain the existence of non-transiting planet, hence adding missing pieces to the architecture of the systems \citep{Xie2014,Zhu2018}, allowing for a better comparison with synthetic planetary system population synthesis \citep[see for example][]{Mordasini2009, Alibert2013,Mordasini2018,Coleman2019,Emsenhuber2020}. TTVs can also be used to constrain the masses of the planets involved \citep[see for example][]{Nesvorny2013}, hence their density, which ultimately give constraints on their internal structures, as is the case for the Trappist-1 system \citep{Grimm2018,Agol2020}. Detection of individual dynamically active systems also provides valuable constraints on planetary system formation theory, as the current orbital state of a system can display markers of its evolution \citep[see for example][]{BaMo2013,Delisle2017}. Orbital interactions also impact the possible rotation state of the planets \citep{DeCoLeRo2017}, hence their atmosphere \citep{Leconte2015}. 
 \begin{figure*}[!ht]
\begin{center}
\includegraphics[width=0.49\textwidth]{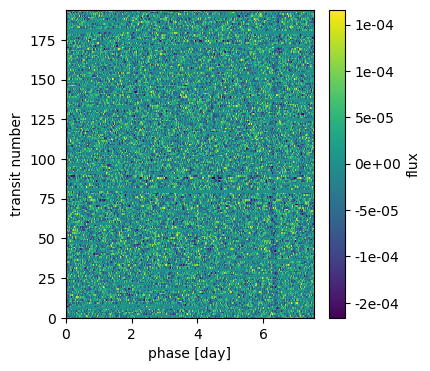}\includegraphics[width=0.49\textwidth]{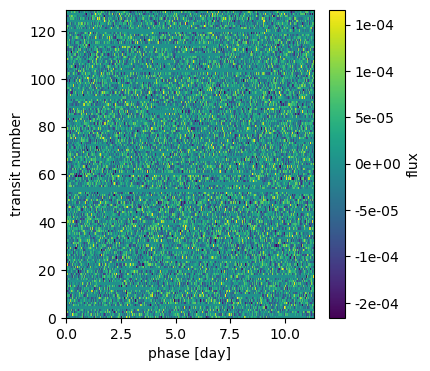}\\
\includegraphics[width=0.49\textwidth]{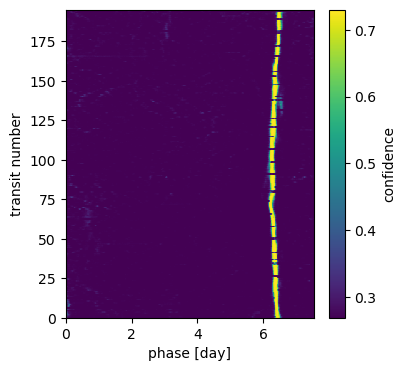}\includegraphics[width=0.49\textwidth]{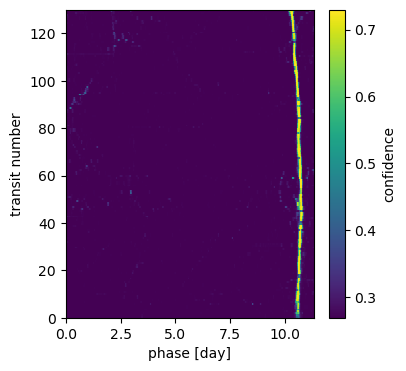}\\
\caption{\label{fig:RIVERS7p5} \textit{Top:} Rivers diagrams of KOI3184 at the period 7.5472d (\textit{left}) and 11.3251d (\textit{right}). The bottom row displays the first 7.5472 (resp. 11.3251) days of data for KOI3184, the color code representing the normalised flux. Each subsequent raw display a new set of  7.5472 (resp. 11.3251) days of data. The flux has been clipped at $3\sigma$ for visibility, and missing data have been replaced by a flux of 1.
\textit{Bottom:} corresponding RIVERS.deep confidence matrices, which shows for each timing of the rivers diagram the confidence for it to belong to the track of a planet. See section \ref{sec:RIVERSdeep} for more details about the confidence matrix.}
\end{center}
\end{figure*}

However, TTVs can also be a bias that affects negatively the detection and characterisation of exoplanets. As previously stated, transit surveys rely on stacking the lightcurve over constant period to extract the shallow transits from the noise. If TTVs of amplitude comparable to - or greater than - the duration of the transit occur on a timescale comparable to - or shorter than - the mission duration, there is not a unique period that will successfully stack the transits of the planet \citep{Garcia2011}. 
This can lead to two problems: incorrect estimates of the planet parameters, and/or the absence of detection. To alleviate this bias, we developed the RIVERS method, based on the recognition of the tracks of planets in river diagrams using machine learning, and a photo-dynamic fit of the lightcurve. The method is described in details in \cite{RIVERS1}.

In this paper we apply the RIVERS method to KIC 4725681, which have 3 KOIs announced on the Kepler database\footnote{\href{https://exoplanetarchive.ipac.caltech.edu/cgi-bin/TblView/nph-tblView?app=ExoTbls\&config=cumulative}{\url{https://exoplanetarchive.ipac.caltech.edu/cgi-bin/TblView/nph-tblView?app=ExoTbls\&config=cumulative}}}, the candidates KOI3184.01 and .03 with orbital periods of 7.54 and 4.02 days, respectively, and the False-positive KOI3184.02 at a period of 11.32 day, flagged as 'Not transit-like'.

\section{Detection of Kepler-1972}
\label{sec:detection}

\subsection{Application of the RIVERS.deep method}

\subsubsection{Preparation of the lightcurve}
\label{sec:lc}
The raw PDCSAP flux is downloaded using the {\ttfamily lightkurve}\footnote{https://docs.lightkurve.org/} package. We started by removing long-term trends of the flux using the \textit{flatten}\footnote{https://docs.lightkurve.org/reference/api/lightkurve.LightCurve.flatten.html} method of the {\ttfamily lightkurve} package using the default parameters. This method apply a Savitzky-Golay filter on the lightcurve. We then checked for gaps longer than 2.5 hours. Such gaps were commonly produced by the monthly data downlinks. After repointing the spacecraft, there was usually a photometric offset produced due to thermal changes in the telescope. We hence removed all data points until the average flux over 4 hours is within 1 standard deviation of the median flux of the lightcurve. Finally, we removed points at more that 4 times the standard deviation from the median value of the lightcurve to remove outliers.

\subsubsection{Application of RIVERS.deep}
\label{sec:RIVERSdeep}

 \begin{figure}[!ht]
\begin{center}
\includegraphics[width=0.49\textwidth]{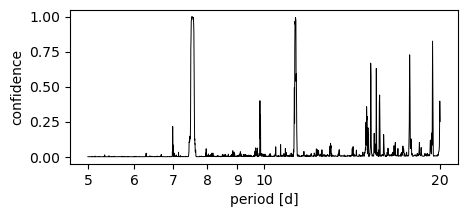}
\caption{\label{fig:periodo} RIVERS periodogram of KIC 4735826: For each period we show the confidence of the RIVERS.deep classifier that the corresponding river diagram contain the track of a planet. {Since the false positive rate of the classifier is typically of a few percent, a photo dynamical fit of the data is required to confirm the planetary nature of the signals.}}
\end{center}
\end{figure}

The RIVERS.deep method, introduced in details in \cite{RIVERS1}, is based on the recognition of the track of a planet in a river diagram \citep{Carter2012}. An example of such diagram is given on the top panel of Fig. \ref{fig:RIVERS7p5}. The RIVERS.deep algorithm takes as input this 2D array and produces two outputs:
\begin{itemize}
\item{\bf a confidence matrix:} an array of the same size as the input containing for each pixel the confidence that this pixel belongs to a transit. This task is performed by the 'semantic segmentation' (pixel-level vetting) subnetwork \citep{jegou2017one}. 
\item{\bf a global prediction:} a value between 0 and 1 which quantifies the model confidence that the output of the semantic segmentation module is due to the presence of a planet. This task is performed by the classification subnetwork.
\end{itemize}

\begin{figure*}[!ht]
\begin{center}
\includegraphics[width=0.99\textwidth]{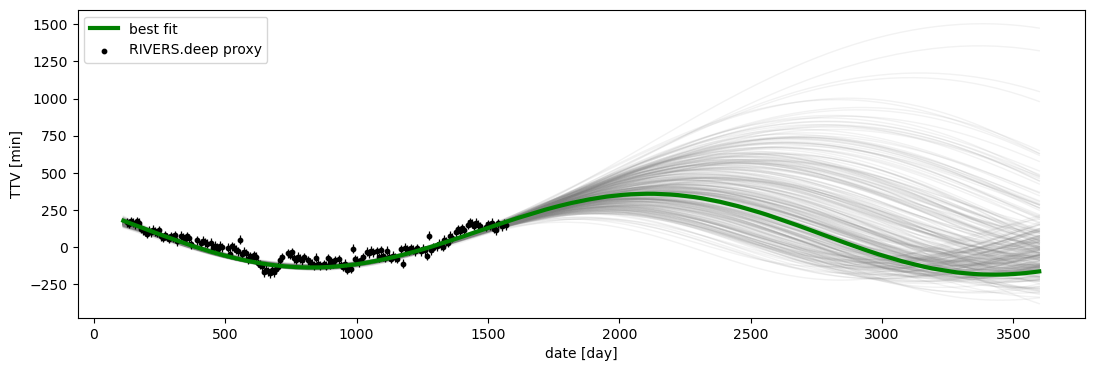}\\
\includegraphics[width=0.99\textwidth]{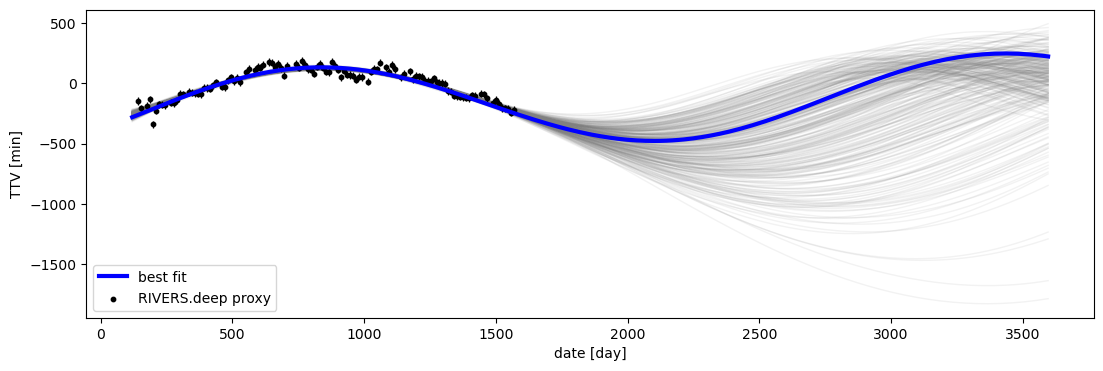}
\caption{\label{fig:KOI3184_TTV} TTVs for Kepler-1972 b (\textit{top}) and Kepler-1972 c (\textit{bottom}). The black errorbars represent the TTVs coming from the RIVERS.deep method {which is the network's highest confidence for the timing of each transit (highlighted pixels in Fig. \ref{fig:RIVERS7p5}). These timings are used as TTV proxy to initialise the photodynamical fit close to its final solution, but are not transit timings fitted to the data. Their errorbar indicate the sensitivity of the method (30 min)}. In grey are 300 samples resulting from the photodynamic fit of the lightcurve, the solid coloured curves correspond to the best fit.}
\end{center}
\end{figure*}

An example of confidence matrix is shown on the bottom panel of Fig. \ref{fig:RIVERS7p5}. The pixels recognized as belonging to a transit are highlighted in yellow. As described in section 3.2.3 of \cite{RIVERS1}, a periodogram can be obtained by saving the output of the global prediction of the model over a series of river diagrams, made for a grid of folding periods. The RIVERS.deep periodogram of KIC 4735826 is displayed in Fig. \ref{fig:periodo}. This periodogram shows two strong peaks at 7.5d and 11.3d. One of the river diagram belonging to the 7.5d (resp. 11.3d) peak is the top-left (resp. top-right) diagram in Fig. \ref{fig:RIVERS7p5}. The transit timings proxy highlighted in the bottom panels of Fig. \ref{fig:RIVERS7p5} are recovered and displayed as black data points in Fig. \ref{fig:KOI3184_TTV}.
{We emphasise that
the word `confidence' used here refer to the output of a neural network, which is not a likelihood or probability. As such, the application of RIVERS.deep is only a preliminary step to the photo dynamical fit of the lightcurve presented in the next section.}

\subsection{Planet detection} 
\label{sec:fit}

\subsubsection{Stellar properties}

The luminosity of KIC 4735826 is computed from the {\it Gaia} Early Data Release 3 parallax \citep{gai21} by taking into account the correction {to the parallax} from \citet{lin21}. The $K$ magnitude of the star is used together with the bolometric correction of \citet{cas14,cas18} and the extinction from the dust map of \citet{gre18}. We then obtain a luminosity $L=1.97 \pm 0.07$ \,$L_{\odot}$ for KIC 4735826. The effective temperature and metallicity of KIC 4735826 are taken from the spectroscopic analysis of \citet{bre18}: $T_{\mathrm{eff}}=5818 \pm 27$\,K and [M/H$]=0.23 \pm 0.01$\,dex. 

These observational constraints ($L$, $T_{\mathrm{eff}}$ and [M/H]) are then used to determine the global parameters of KIC 4735826 from stellar models computed with the Geneva stellar evolution code \citep{egg08}. For this determination, the adopted uncertainty on the effective temperature was increased to 50\,K (instead of the small internal error of 27\,K) to better account for possible uncertainties coming from different spectroscopic determinations. We then find that KIC 4735826 is a star at the end of its evolution on the main sequence with a mass of $M=1.12 \pm 0.03$\,$M_{\odot}$, a radius of $R=1.384 \pm 0.050$\,$R_{\odot}$ and an age of $7.4 \pm 1.2$\,Gyr. The stellar properties are summarized in Table~\ref{tab:models_ge}.

\begin{table} [h!]
\caption{Stellar global properties of Kepler-1972. The upper panel provides the adopted observational constraints used to determine the global stellar properties from stellar models given in the lower panel of the Table.} 
\label{tab:models_ge} 
\label{tab:stellarParam} 
\centering 
\begin{tabular}{lll} 
\hline\hline 
\multicolumn{3}{c}{Kepler-1972}\\ 
KIC & \multicolumn{2}{l}{4735826}\\ 
\hline 
Parameter & Value & Reference \\ 
\hline 
mKep [mag] &11.158 & 1 \\ 
mV [mag] & $11.230\pm0.034$ & 1 \\ 
mJ [mag] & $10.118\pm0.021$ & 1 \\ 
mH [mag] &  $ 9.858\pm0.020$ & 1 \\ 
\hline 
$T_{\mathrm{eff}}$ [K] & $5818 \pm 27 $ & 2 \\ 
 $[$M/H$]$ [dex] & $0.23 \pm 0.01$ & 2 \\
$L_{\star}$ [$L_{\odot}$] &  $1.97 \pm 0.07$ & 3 \\  
\hline 
$M_{\star}$ [$M_{\odot}$] &  $1.12 \pm 0.03$  & 3 \\ 
$R_{\star}$ [$R_{\odot}$] &  $1.384 \pm 0.050$  & 3 \\
$t_{\star}$ [Gyr]  &  $7.4 \pm 1.2$  & 3 \\ 
$\log{g}$ [cgs]      &  $4.20 \pm 0.04$ & 3 \\
$\rho_{\star}$ [$\rho_\odot$] &  $0.422 \pm 0.050$  & 3 \\
\hline 
\end{tabular} 
\tablefoot{  
[1] https://exoplanetarchive.ipac.caltech.edu  [2] \cite{bre18}, [3] this study } 
\end{table}


\subsubsection{Planetary solution}

\begin{table*} 
\begin{small} 
\caption{Fitted and derived properties of the planets and candidate of the Kepler-1972 system} 
\label{tab:planetParam} 
\centering 
\begin{tabular}{lllll} 
\hline 
Parameter &  Prior & Keplerian & n-body  & \\ 
\hline\hline 
\multicolumn{5}{c}{KOI3184.03}\\ 
\hline 
$P$ [day]&U[8,12]&$4.020192_{-3.4e-05}^{+3.5e-05}$&$4.020184_{-3.5e-05}^{+3.6e-05}$&fitted\\ 
$R/R_\star$ &U[0,1e-1]&$0.00372_{-1.9e-04}^{+1.9e-04}$&$0.00365_{-1.8e-04}^{+1.8e-04}$&fitted\\ 
$b$&U[0,1]&$0.51_{-0.11}^{+0.08}$&$0.39_{-0.18}^{+0.13}$&fitted\\ 
$t0$ [BJD-2454833.0]& &$134.8402_{-0.0080}^{+0.0083}$&$134.8424_{-0.0082}^{+0.0082}$&fitted\\ 
$a/R_\star$ & &$7.98_{-0.32}^{+0.30}$&$8.11_{-0.29}^{+0.27}$&derived\\ 
$I$ [deg]& &$86.35_{-0.58}^{+0.77}$&$87.25_{-0.97}^{+1.31}$&derived\\ 
$R$ [R$_{Earth}$]& &$0.561_{-0.035}^{+0.035}$&$0.551_{-0.034}^{+0.034}$&derived\\ 
SNR& &$11.14$&$11.10$& derived \\ 
\hline\hline 
\multicolumn{5}{c}{Kepler-1972 b (KOI3184.01)}\\ 
\hline 
$\lambda$ [deg]&U[0,360]&$-$&$23.46_{-4.72}^{+4.79}$&fitted\\ 
$P$ [day]&U[1,20]&$7.54765_{-1.3e-04}^{+1.2e-04}$&$7.54425_{-5.4e-04}^{+5.4e-04}$&fitted\\ 
$e\cos \varpi$ &U[-.9,.9]$^*$&$-$&$0.014_{-0.043}^{+0.045}$&fitted\\ 
$e\sin \varpi$ &U[-.9,.9]$^*$&$-$&$0.045_{-0.044}^{+0.060}$&fitted\\ 
$M_{pl}/M_\star$ &U[0,1e-2]&$-$&$5.4e-06_{-1.7e-06}^{+1.5e-06}$&derived\\ 
$R_{pl}/R_\star$ &U[0,1e-1]&$0.00391_{-1.6e-04}^{+1.6e-04}$&$0.00532_{-1.9e-04}^{+2.0e-04}$&fitted\\ 
$b$&U[0,1]&$0.13_{-0.09}^{+0.13}$&$0.51_{-0.15}^{+0.09}$&fitted\\ 
$t0$ [BJD-2454833.0]& &$134.406_{-0.013}^{+0.016}$&$111.8999_{-0.0090}^{+0.0078}$&f/d\\ 
$a/R_\star$ & &$12.15_{-0.49}^{+0.45}$&$12.34_{-0.44}^{+0.41}$&derived\\ 
$e$ & &$-$&$0.067_{-0.040}^{+0.071}$&derived\\ 
$\varpi$ [deg]& &$-$&$64.9_{-57.2}^{+50.1}$&derived\\ 
$I$ [deg]& &$89.05_{-0.94}^{+0.67}$&$87.62_{-0.49}^{+0.73}$&derived\\ 
$M_{pl}$ [M$_{Earth}$]& &$-$&$2.02_{-0.62}^{+0.56}$&derived\\ 
$R_{pl}$ [R$_{Earth}$]& &$0.590_{-0.033}^{+0.032}$&$0.802_{-0.041}^{+0.042}$&derived\\ 
$\rho_{pl}$ [$\rho_{Earth}$]& &$-$&$4.02_{-1.31}^{+1.44}$&derived\\ 
SNR& &$12.38$&$18.96$& derived \\ 
\hline\hline 
\multicolumn{5}{c}{Kepler-1972 c  (KOI3184.02)}\\ 
\hline 
$\lambda$ [deg]&U[0,360]&$-$&$156.60_{-3.57}^{+4.06}$&fitted\\ 
$P$ [day]&U[1,20]&$11.32057_{-2.5e-04}^{+2.8e-04}$&$11.3295_{-0.0011}^{+0.0011}$&fitted\\ 
$e\cos \varpi$ &U[-.9,.9]$^*$&$-$&$-0.007_{-0.037}^{+0.031}$&fitted\\ 
$e\sin \varpi$ &U[-.9,.9]$^*$&$-$&$-9.6e-04_{-0.036}^{+0.046}$&fitted\\ 
$M_{pl}/M_\star$ &U[0,1e-2]&$-$&$5.7e-06_{-1.7e-06}^{+1.6e-06}$&derived\\ 
$R_{pl}/R_\star$ &U[0,1e-1]&$0.00398_{-3.3e-04}^{+3.2e-04}$&$0.00575_{-2.6e-04}^{+2.6e-04}$&fitted\\ 
$b$&U[0,1]&$0.68_{-0.11}^{+0.07}$&$0.810_{-0.031}^{+0.033}$&fitted\\ 
$t0$ [BJD-2454833.0]& &$142.839_{-0.020}^{+0.019}$&$119.800_{-0.012}^{+0.014}$&f/d\\ 
$a/R_\star$ & &$15.91_{-0.64}^{+0.59}$&$16.19_{-0.58}^{+0.54}$&derived\\ 
$e$ & &$-$&$0.043_{-0.028}^{+0.046}$&derived\\ 
$\varpi$ [deg]& &$-$&$-10_{-119}^{+139}$&derived\\ 
$I$ [deg]& &$85.13_{-0.58}^{+0.80}$&$87.13_{-0.19}^{+0.17}$&derived\\ 
$M_{pl}$ [M$_{Earth}$]& &$-$&$2.11_{-0.65}^{+0.59}$&derived\\ 
$R_{pl}$ [R$_{Earth}$]& &$0.601_{-0.055}^{+0.054}$&$0.868_{-0.050}^{+0.051}$&derived\\ 
$\rho_{pl}$ [$\rho_{Earth}$]& &$-$&$3.33_{-1.07}^{+1.17}$&derived\\ 
SNR& &$7.47$&$12.60$& derived \\ 
\hline\hline 
\multicolumn{5}{c}{Kepler-1972 b and c}\\ 
\hline 
$log_{10}((m_{b}+m_{c})/m_\star)$&U[-7,-2]&$-$&$-4.96_{-0.16}^{+0.11}$&fitted\\ 
$m_{b}/(m_{b}+m_{c})$&U[0,1]&$-$&$0.489_{-0.014}^{+0.014}$&fitted\\ 
\hline\hline 
\multicolumn{5}{c}{Kepler-1972}\\ 
\hline 
$\rho_{\star}$ [$\rho_\odot$]&$\mathcal N$(0.422,0.050)&$0.423_{-0.049}^{+0.049}$&$0.444_{-0.046}^{+0.046}$&fitted\\ 
limbdark $u_1$&$\mathcal N$(0.454,0.026)&$0.455_{-0.026}^{+0.026}$&$0.454_{-0.025}^{+0.025}$&fitted\\ 
limbdark $u_2$&$\mathcal N$(0.236,0.036)&$0.237_{-0.036}^{+0.036}$&$0.237_{-0.036}^{+0.035}$&fitted\\ 
log10(jitter)&U[-10,0]&$-4.3931_{-0.0022}^{+0.0022}$&$-4.3951_{-0.0022}^{+0.0022}$&fitted\\ 
$log10(\sigma_{GP})$ &U[-10,0]&$-8.27_{-1.16}^{+1.16}$&$-8.24_{-1.16}^{+1.10}$&fitted\\ 
$log10(\tau_{GP})$ [day] &U[0,3]&$1.50_{-1.02}^{+1.01}$&$1.51_{-1.00}^{+1.00}$&fitted\\ 
\end{tabular} 
\tablefoot{  
{Orbital elements are given a the date 110.5394 [BJD-2454833.0]. The main value is the median of the posterior distribution. The reported errors are the distance to the 0.15865 quantile (lower error) and 0.84135 quantile (upper error). U refer to a flat prior between the given bounds, while $\mathcal N(\mu,\sigma)$ refer to a graussian distribution of mean $\mu$ and standard deviation $\sigma$. U$^*$: For the $k=e\cos \varpi$ and $h=e\sin \varpi$ variables, an additional prior was added to enforce an uniform distribution for their module $e$. f/d indicates that the variable is fitted for the circular model, but derived in the n-body model (where $\lambda_j$ is the fitted parameter). The reported SNR is the median depth of transit over its standard deviation to better account for potential degeneracy with the noise modelling at the transit location.}}
\end{small} 
\end{table*}

 \begin{figure*}[!ht]
\begin{center}
\includegraphics[width=0.49\textwidth]{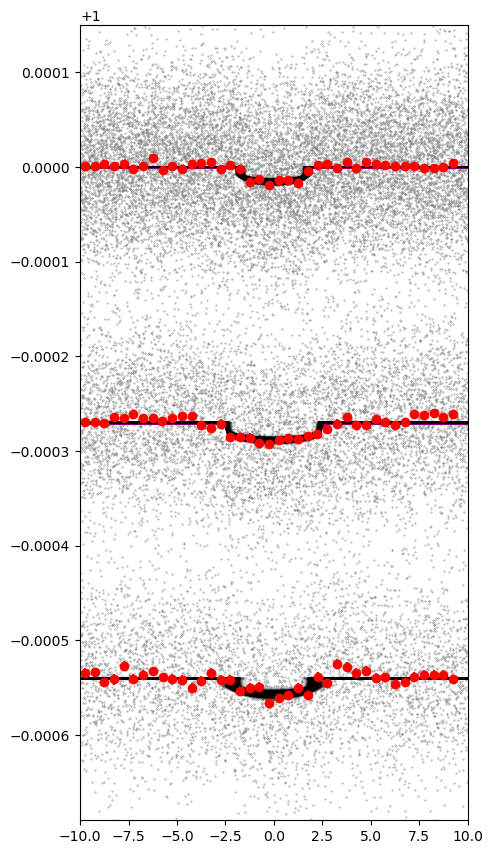}
\includegraphics[width=0.49\textwidth]{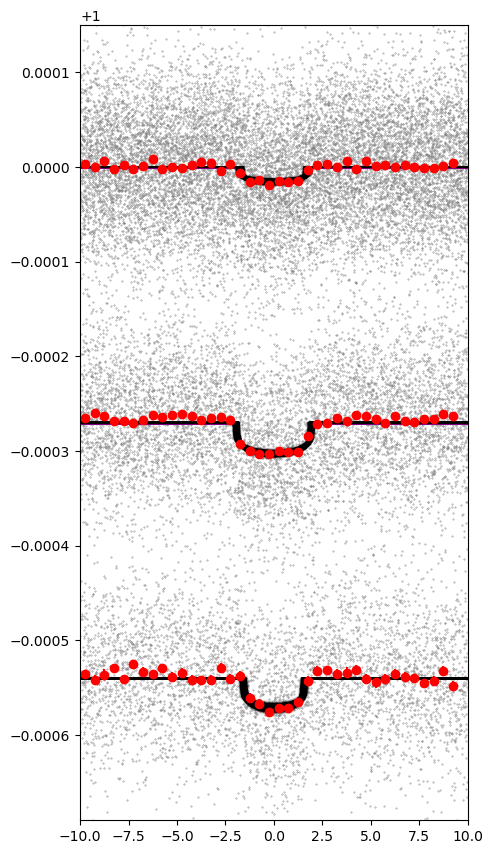}
\caption{\label{fig:KOI3184stacked} Stacked transits of KOI3184.03 (\textit{top}),  Kepler-1972 b (\textit{middle}), and Kepler-1972 c (\textit{bottom}). The left panel results from the stacking along the Keplerian ephemeride, which is the left solution in table \ref{tab:planetParam}.  The right panel result from the stacking along the Keplerian ephemeride, for the inner planet, and the n-body solution for Kepler-1972 b and c, which is the right solution in table \ref{tab:planetParam}}
\end{center}
\end{figure*}

The fit of the lightcurve was performed with the same setup as the one presented in \cite{RIVERS1}. To highlight the effect of the TTVs on the recovered transit parameters, we performed two distinct fits of the lightcurve: a fit with keplerian orbits (constant periods) for the three candidates at 4, 7.5, and 11.3 days, and a photodynamic fit where the transit timings of the two resonant planets where modelled using the {\ttfamily TTVfast} algorithm \citep{DeAgHo2014}. The approximate initial conditions for the orbital elements and masses of these planets were obtained by a preliminary fit of the transit timings to the timing proxy shown in Fig. \ref{fig:KOI3184_TTV}. In the photodynamic fit, the orbit of the 4d candidate is still considered as Keplerian since its interactions with the other two planets are negligible over the duration of the Kepler mission. 

For both fits, {we use the adaptive MCMC sampler samsam\footnote{\url{https://gitlab.unige.ch/Jean-Baptiste.Delisle/samsam}} \citep[see][]{Delisle2018}, which learns the covariance of the target distribution from previous samples in order to improve the subsequent sampling efficiency.}
%
{The likelihood is defined as :
\be
\begin{aligned}
\ln \mathcal L (\theta,\alpha) =& \ln p(y|\theta,\alpha)\\
=&- \frac{1}{2}\left(y-m(\theta)\right)^TC^{-1}(\alpha)\left(y-m(\theta)\right)\\
&- \frac{1}{2} \ln \det \left( 2 \pi C (\alpha) \right)
\end{aligned}
\ee
where $y$ is the photometric data and $\theta$ is the vector of the fitted planetary parameters given in table \ref{tab:planetParam} as well as the limb-darkening coefficients and stellar density. 
The lightcurve model $m(\theta)$ was obtained by computing the transit timings for the chosen type of orbit (Keplerian ephemerides or n-body), then by modeling the transits of each planet with the {\ttfamily batman} package \citep{batman}, with a supersampling parameter set to $29.42$ minutes to account for the long exposure of the dataset. 
The effective temperature, ${\log g}$ and metallicity of the star (Table \ref{tab:stellarParam}) were used to compute the quadratic limb-darkening coefficients $u_1$ and $u_2$ and their error-bars adapted to the Kepler spacecraft using {\ttfamily LDCU}\footnote{https://github.com/delinea/LDCU}. Based on the {\ttfamily limb-darkening} package \citep{EsJo2015}, LDCU  uses two libraries of stellar atmosphere models ATLAS9 \citep{Kurucz1979} and PHOENIX \citep{Husser2013} to compute stellar intensity profiles for a given pass-band. 
Remaining long-term trends were modeled using
the \textsc{s+leaf} \citep{Delisle2020}. \textsc{s+leaf}
 is a C library with python wrappers that implements an optimized GP framework.
While the computational cost of classical GP implementations typically scales has the cube of the dataset size,
the cost of a S+LEAF GP scales linearly \citep{Foreman-Mackey2017,Delisle2020,Delisle2022}.
We used a gaussian process framework with a Matérn 3/2 kernel whose timescale was forced to be above one day (uniform prior of the log of the timescale $\tau_{GP}$ set to U$[0,3]$) to avoid interfering with the modeled transits. A jitter term was also added to all photometric measurements. 
$\alpha$ is the vector of the noise parameters (jitter, $\sigma_{GP}$, $\tau_{GP}$). $C (\alpha)$ is the corresponding covariance matrix.}

.    

The posterior of the fits are summarised in Table \ref{tab:planetParam}, with the Keplerian model on the left and the n-body model on the right. The recovered stacked transits are shown on Fig. \ref{fig:KOI3184stacked}. The difference in the models for the two outer candidates did not affect much the fit of the inner candidate, for which the fit converged to a radius of $0.551_{-0.034}^{+0.034}$ R$_{Earth}$.
The difference is significant for the two resonant planets : the transits stacked along a Keplerian ephemerides yield a strong dispersion of points in and near-transit, responsible for the false positive flag attributed to KOI3184.02 by the Kepler pipeline. Stacking the transits along the TTV-corrected transit timings greatly reduce the in-transit scattering, as can be seen in the right-hand side of Fig. \ref{fig:KOI3184stacked}. The photodynamic fit yields radius of $0.802_{-0.041}^{+0.042}$ and $0.868_{-0.050}^{+0.051}$ R$_{Earth}$ for Kepler-1972 b and c, respectively, while the Keplerian model yields radii smaller by $\sim 28 \%$ and $\sim 35 \%$, respectively. Finally, the photodynamic fit yields projected inclinations of $87.25_{-0.97}^{+1.31}$, $87.62_{-0.49}^{+0.73}$ and $87.13_{-0.19}^{+0.17}$, which are consistent with a coplanar system.

The estimated masses are of $2.02_{-0.62}^{+0.56}$ and $2.11_{-0.65}^{+0.59}$M$_{Earth}$ for Kepler-1972 b and c, respectively, resulting in densities of $4.02_{-1.31}^{+1.44}$ and $3.33_{-1.07}^{+1.17}$ $\rho_{Earth}$. Although the densities are poorly constrained, the planets appear to be denser than the Earth. The mass distribution between the two planets is well constrained $m_{01}/(m_{01}+m_{02})=0.489_{-0.013}^{+0.014}$, while the total mass of the planets is not. This is due to the TTV signal dependency on the parameters: the mass distribution is linked to the relative amplitude of the TTVs between the two planets, while the total planetary mass is constrained by the TTV period \citep{Agol2005,NeVo2016}. The latter could not be properly estimated with the available baseline: as can be seen on the posterior of the TTVs shown in Fig. \ref{fig:KOI3184_TTV}, there is a range of TTV periods that are consistent with the observed signal.

\section{The resonant pair of Kepler-1972}
\label{sec:dyn}

In this section we study a sub-population of 2300 randomly-chosen samples from the posteriors presented in Table \ref{tab:planetParam}. Fig.~\ref{fig:KOI3184stab} shows the projection of these samples in the $(\sqrt{e_1^2+c^2e_2^2},\Delta \varpi = \varpi_2-\varpi_1)$ plane \citep[$c=-1.22$ for the 3:2 MMR and is discussed in section 5.2 of][]{RIVERS1}. The $e_j$ and $\varpi_j$ represented here are the fitted initial conditions at the date 2454943.5394 BJD. 

\subsection{Stability}  

 \begin{figure}[!ht]
\begin{center}
\includegraphics[width=0.49\textwidth]{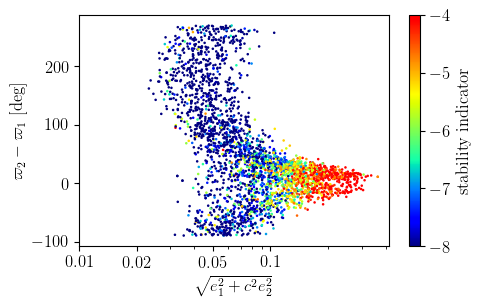}
\includegraphics[width=0.49\textwidth]{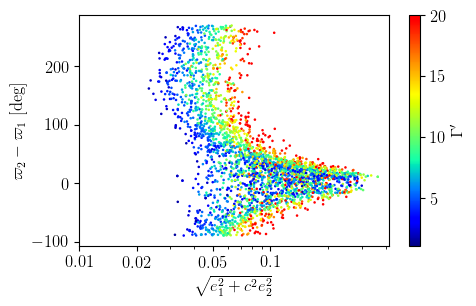}
\includegraphics[width=0.49\textwidth]{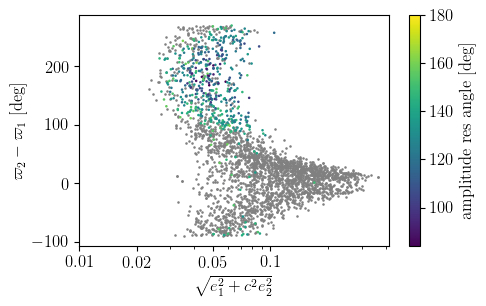}
\caption{\label{fig:KOI3184stab} Posterior of the solution shown in Table \ref{tab:planetParam}. The $c=-1.22$ parameter is discussed in \citep{RIVERS1}.  \textit{Top:} stability of the posteriors, ranging from stable (blue) to unstable (red).
\textit{Middle:} Value of the Hamiltonian parameter $\Gamma'$ over the posterior.
\textit{Bottom:} Smallest amplitude between the two resonant angles. Grey dots indicate a circulation of both of them.}
\end{center}
\end{figure}

We verified the stability of the posteriors using the frequency analysis criterion \citep{La90,La93}, using the same implementation as in \cite{Leleu2021,RIVERS1}. For this stability analysis, the resonant pair and the inner candidate at $4.02$~d were considered. Since only the radius of this inner candidate could be derived in our analysis, we assumed its density to be of $2.6 \rho_{Earth}$, an arbitrary value which is consistent with the densities of the two outer planets. The {top} panel of Fig. \ref{fig:KOI3184stab} shows the resulting criterion for each initial condition, for {n-body} integration over $10^5$ years. We find that the bulk of the low-eccentricity part of the posterior is stable for more than $10^6$ year, which together correspond to more than 45 billion orbits of the resonant pair. On the other hand, a significant part of the posterior is unstable on a short time-scale (red dots on the top panel of Fig. \ref{fig:KOI3184stab}) for eccentricities typically higher than $\sim 0.1$. 

\subsection{Dynamics and TTV degeneracy}

\subsubsection{Resonant state of the system}

 \begin{figure}[!ht]
\begin{center}
\includegraphics[width=0.49\textwidth]{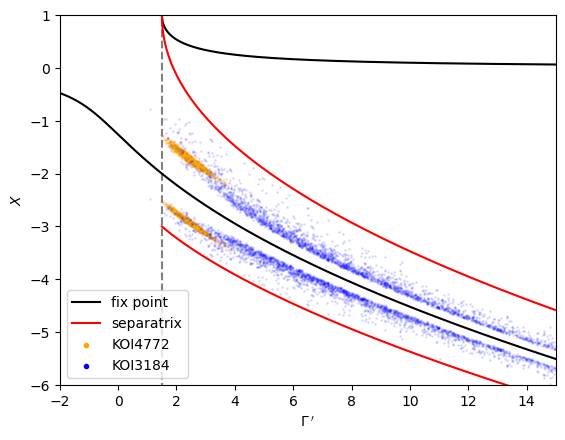}
\caption{\label{fig:XGamma} Surface of section in the ($\Gamma'$,$X$) plane of the Hamiltonian (\ref{eq:ham}). The black line shows the positions of the family of elliptic fixed point (the hyperbolic family exist for larger value of X and is not displayed here). The red line show the intersection of the separatrix with the $Y=0$ plane for each value of $\Gamma '$. The coloured dots show the intersection of the trajectories emanating from the initial conditions of the posterior summarised in section \ref{tab:planetParam} and the ($\Gamma'$,$X$) plane, for Kepler-1972 (this paper) and Kepler-1705 \citep{RIVERS1} }
\end{center}
\end{figure}

We consider the Hamiltonian formulation of the Second Fundamental Model for Resonance of \citep{HenLe1983}. Using the 1-degree of freedom model of first order resonances presented in \cite{DePaHo2013}, the dynamics of a resonant system can be derived from the equations canonically associated with the conjugated variables X and Y of the Hamiltonian:
\begin{equation}
H=\frac{1}{2}\left[\frac{1}{2}\left(X^2+Y^2\right)-\Gamma'\right]^2-X
\label{eq:ham}
\end{equation}
The change of coordinates from the orbital elements and masses to the variables $X$ and $Y$ are given in section 2 of \cite{DePaHo2013}. For $\Gamma' < 1.5$ the model has a single family of elliptic fixed points and no separatrix. At $\Gamma' =1.5$ a bifurcation occurs and for $\Gamma' >1.5$ there are two families of elliptic fix points and one family of hyperbolic fix points from which emanates a separatrix. Systems are considered formally resonant when $\Gamma' >1.5$ and they lie inside the separatrix. Since $Y=0$ for all fixed points, and all trajectories of the Hamiltonian \ref{eq:ham} cross the $Y=0$ line, we show a surface of section of the Hamiltonian for $Y=0$ in Fig. \ref{fig:XGamma}. This representation is equivalent to the one presented in \cite{NeVo2016,Nesvorny2021}. The blue dots represent the posterior of Kepler-1972, while the orange dots show the position of Kepler-1705, for comparison. Despite the large uncertainties on the orbital parameters and masses due to the observation baseline shorter than the resonant period, the whole posterior lies inside the 3:2 MMR. 

\subsubsection{TTV degeneracy}

In \cite{RIVERS1} we introduced the variables 
\be
u=e_b\,{\rm e}^{i\phi_1}+c e_c\,{\rm e}^{i\phi_2} 
\hspace{0.5cm}{\rm and}\hspace{0.5cm}
v=ce_b\,{\rm e}^{i\phi_1}-g e_c\,{\rm e}^{i\phi_2}
\label{eq:uv}
\ee
(Mardling, in prep), where
\be
\phi_1= 2 \lambda_{b}-3 \lambda_{c} + \varpi_{b} 
\hspace{0.5cm}{\rm and}\hspace{0.5cm}
\phi_2= 2 \lambda_{b}-3 \lambda_{c} + \varpi_{c} 
\label{resangs}
\ee
are the resonant angles for the 3:2 resonance,
$c$ is a function of the Laplace coefficients whose value is $-1.23$ for the 3:2 commensurability, and $g=(m_2/m_1)\sqrt{a_2/a_1}$. For Kepler-1705 \citep[the system discovered in][]{RIVERS1}, a full resonant period is observed, and as a result $\Gamma '$ (and the corresponding range of values of $u$) is well determined and varies little across the posterior. In that case, we showed that the theoretical posterior of the variables $e_j$ and $\varpi_j$ could be obtained by varying the real and imaginary parts of $v$ in the range $[-0.2,0.2]$; indeed, the TTV signal constrains the resonant part of the eccentricity $u$, but is blind to the free part  $v$ \citep[Mardling, in prep,][]{RIVERS1}. In Kepler-1972 however, $\Gamma'$ (and $u$) is poorly constrained across the posterior, which parametrizes an additional degeneracy for the $e_j$ and $\varpi_j$ variables; see the middle panel of Fig.\ref{fig:KOI3184stab}.



\section{Summary and Conclusion}
\label{sec:conclusion}

For planets that are too small to induce individually-detectable transits, the shape of their transit is derived from light curves that are stacked along constant periods. In this case, TTVs can lead to erroneous estimations of the transit depth and duration, or even create a signal that is not recognised as a transit anymore. 
We illustrate this on KOI-3184.02, which is flagged as 'Not transit-like false positive' on the NASA Exoplanet Archive as of November 2021. Applying the RIVERS methods, we retrieved the track of KOI-3184.01 and KOI-3184.02 in the lightcurve. We show that both of these planets have anti-correlated TTVs of 6 hour of peak-to-peak amplitude, these TTVs being responsible for the apparent incoherence of their transit signature. We now name these planets  Kepler-1972 b and c. As for Kepler-1705 \citep{RIVERS1}, the recovered planets have individual transit SNR of $\sim 1$, showcasing the performance of the approach to recover individual transits that would otherwise be lost in the noise.  

Kepler-1972 b and c is a pair of Earth-sized planets with $R_b=0.802_{-0.041}^{+0.042}R_{Earth}$ and $R_c=0.868_{-0.050}^{+0.051}R_{Earth}$, with similar masses $m_b/m_c =0.956_{-0.051}^{+0.056}$. The observed TTV signal is enough to show that the pair is formally inside the 3:2 mean motion resonance (inside the separatrix). However, the baseline of the observation is not long enough to cover a full period of the TTV signal, leading to a somewhat imprecise estimation of the planet masses, putting an $1\sigma$ upper limit at $M_b=2.58 M_{Earth}$ and $M_c=2.70 M_{Earth}$. {Fitting the inner candidate (KOI-3184.03, $P= 4.02$[day]) at the same time as the resonant pair, we show that the three projected inclinations are 1$\sigma$ compatible: $I_{.03}=87.25_{-0.97}^{+1.31}$,
$I_{b}=87.62_{-0.49}^{+0.73}$, and
$I_{c}=87.13_{-0.19}^{+0.17}$ degrees. This is consistent with a co-planar 3-planet system, which increases the probability that the 4.02d signal is also of planetary nature. }

Recovering a planetary signal from a Kepler false positive shows that special care needs to be taken in the treatment of low-S/R planet candidates in transit surveys, as the stacking methods broadly used is not well suited when planet-planet interactions induce TTVs larger than the transit duration. This effect is especially relevant for the Kepler mission, TESS polar observations, and the upcoming PLATO mission.

\begin{acknowledgements}
This work has been carried out within the framework of the National Centre of Competence in Research PlanetS supported by the Swiss National Science Foundation and benefited from the seed-funding program of the Technology Platform of PlanetS. The authors acknowledge the financial support of the SNSF.
\end{acknowledgements}

\bibliographystyle{aa}
\bibliography{biblio_RIVERS_deep}

\appendix

\end{document}